\documentclass[twocolumn,pra,aps]{revtex4}
\usepackage{graphicx}
\usepackage{amsmath}
\usepackage{amssymb}
\usepackage{centernot}
\usepackage{enumerate}

\begin{document}

\title{All-Possible-Worlds:\\
Unifying Many-Worlds and Copenhagen, in the Light of Quantum Contextuality.}

\author{Antoine Suarez}
\affiliation {Center for Quantum Philosophy \\ Ackermannstrasse 25, 8044 Z\"{u}rich, Switzerland\\
suarez@leman.ch, www.quantumphil.org}

\date{December 15, 2017-March 4, 2019}

\begin{abstract}

``All-Possible-Worlds" is a novel interpretation of quantum physics, which results from a unified reformulation of Many-Worlds and Copenhagen (``collapse") in the light of quantum contextuality, and proposes ``nonlocality at detection" as a principle ruling the whole quantum realm, including single qubit and qutrit experiments.

\ \\
\textbf{Keywords:} Many-worlds, quantum contextuality, nonlocality, free will, divine omniscience, discreteness of space-time, heat death of the universe.

\end{abstract}


\maketitle

\section {Introduction} \label{intr}

``Fifty years ago, Simon Kochen and Ernst Specker proved that quantum theory cannot be explained with noncontextual models.[...]In the last decade, the Kochen-Specker theorem has inspired some results which help us to understand quantum theory and [...] the power of quantum systems for information processing and computation." \cite{workshop, ks67}.

At the same time, Specker's ``theological motivation" behind the theorem has received little attention. Specker was an inspired theologian seriously engaged in interpreting Scripture, as his insightful sermons in Z\"{u}rich's Evangelische Hoschschulgemeinde and Predigerkirche show \cite{Specker78}. One of the theological questions he worried about is that of divine omniscience. In his seminal article ``Die Logik nicht gleichzeitig entscheindbarer Aussagen" in 1960 Specker brings this question in relation with quantum contextuality \cite{Specker60}:

\begin{footnotesize}
``In a certain sense the scholastic speculations about the ``Infuturabilien" [this term invented by Specker is to be translated as something like`future contingencies'] also belong here,that is, the question whether the omniscience of God also extends to events that would have occurred in case something would have happened that did not happen. (cf. e.g. [3], Vol. 3, p. 363.)"
\end{footnotesize}

In this quotation Reference [3] is the book by M. Solana, Historia de la filosof\'{i}a espa\~{n}ola (Asociacion Espa\~{n}ola para el Progreso de las Ciencias, Madrid 1941), where in ``p. 363" is presented the philosophical view of the Portuguese Jesuit Pedro de Fonseca. And the ``scholastic speculations" Specker refers to is the controversy raised by the theory of Luis de Molina (disciple of Fonseca) about the compatibility of ``Divine omniscience and human free-will".

In this article I argue that the question of ``Infuturabilien" referred to by Specker in \cite{Specker60} may help us to better understand the relationship between quantum contextuality and nonlocality. Specifically, ``infuturabilien" allow us to reformulate ``Many-Worlds" as a theory of ``All-Possible-Worlds", which do exist in the divine mind \cite{aa}: By making a choice to perform a measurement the experimenter realizes one of these possible worlds, instead of provoking the split of the actual world (and the experimenter her/himself) into many unconnected worlds, as Many-Worlds claims.

In this sense All-Possible-Worlds makes it possible to unify two interpretations of quantum physics considered opposite so far: Conpenhagen and Many-Worlds, and proposes ``nonlocality at detection" as a principle ruling the whole quantum realm, including single qubit and qutrit experiments. In the same line of thinking it accounts for divine omniscience without abolishing human free-will, and thereby contributes to solve the outstanding ``Infuturabilien" problem.

Additionally, All-Possible-Worlds has a number of interesting consequences for physics: The number of possible worlds in the divine mind is the same as the maximal number of free choices the humans of all times can in principle perform. If one acknowledges the theological theorem that this number is finite, and asks for physical features granting this, one is naturally lead to: 1) discreetness of space-time, 2) upper bounded signaling, and 3) the heat death of the universe (``time arrow" and ``second law" of thermodynamics).

All this means that what is and is not possible is not determined by physical ``laws" but the other way around, it is these ``laws" which actually arise from what is and is not possible. All-Possible-Worlds bring to focus an overlooked feature of Many-Worlds: At the end of the day physical reality is defined by the free choices human observers can perform.

In summary, reformulating Many-Worlds in the light of Quantum Contextuality allow us to appreciate the astonishing potential of quantum physics for interpreting consistently the physical world without giving up free-will, stimulating interdisciplinary research, and contributing even to philosophical and theological knowledge, the kind of work Ernst Specker did. Paraphrasing Alexei Grinbaum: ``By linking the mathematical properties of [the quantum] with other episodes in the history of ideas, one gives quantum theory and quantum technologies a place in history and a place in culture." \cite{ag17}

\section {The ``Assyrian prophet" parable} \label{parab}
Specker liked to introduce quantum contextuality in a narrative way by using the following parable he invented \cite{Specker60}:

\begin{footnotesize}
``During the age of king Asarhaddon a wise man from Ninive taught at the school of prophets in Araba'ilu. He was an outstanding representative of his discipline (solar and lunar eclipses) who was, except for the heavenly bodies, concerned almost exclusively about
his daughter. His teaching success was modest, the discipline was proved to be dry and required also previous mathematical knowledge, which was scarcely present. Although he did not find the interest amongst the students that he had hoped for, it was however given to him
in abundance in a different field. No sooner had his daughter reached the marriageable age, than he was bombarded with marriage proposals to her by students and young graduates. And although he did not believe that he could keep her to himself forever, she was in any case still far too young and her suitors were also in no way worthy of her.
And in order for each of them to convince themselves that they were unworthy, he promised them that she would be the wife of he who would solve a prediction-task that was posed to them."
\end{footnotesize}

\begin{figure}[t]
\includegraphics[trim = 4mm 150mm 7mm 0mm, clip, width=0.99\columnwidth]{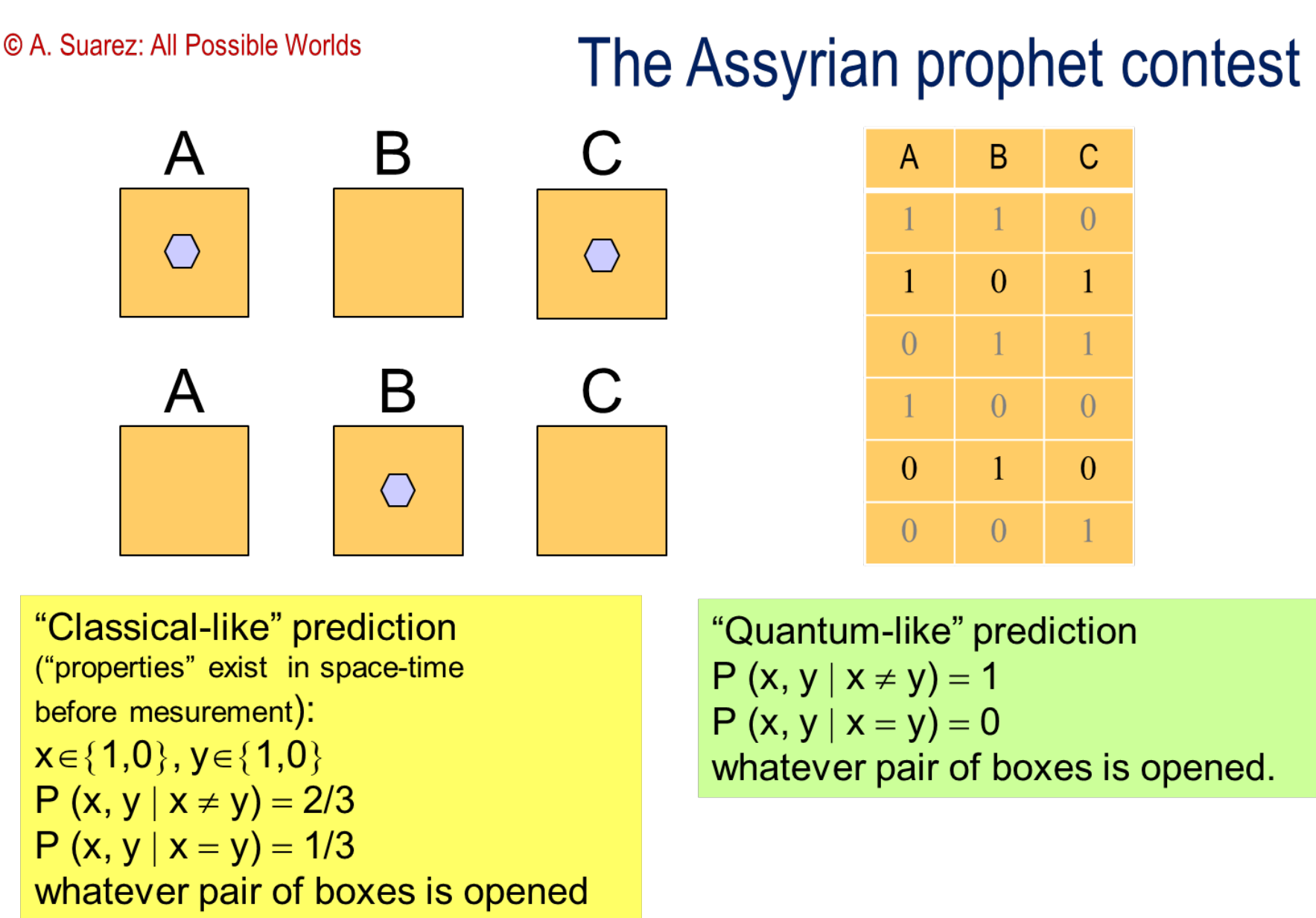}
\caption{\textbf{Quantum Contextuality:} In Ernst Specker's parable (see text) the suitors were led in front of a table on which three
boxes were positioned in a row, and they were ordered to indicate which of the boxes contained
a gem and which were empty. And now no matter how many times they tried, it seemed to be impossible to solve the task. After their predictions, each of the suitors was ordered to open two boxes which they had indicated to be both
empty or both not empty: it turned out each time that one contained a gem and the other did
not, and, to be precise, sometimes the gem was in the first, sometimes in the second of the
boxes that were opened.\cite{Specker60}}
\label{f1}
\end{figure}

The prediction-task is represented in Fig.\ref{f1}. Specker arranged a happy-end to the story as follows \cite{Specker60}:

\begin{footnotesize}
The daughter would have remained unmarried until
the father's death, if she would not have swiftly opened two boxes herself after the prediction
of the son of a prophet, who indicated that precisely one should be filled and the other empty,
which turned out to be actually the case. At the weak objection by the father that he would
have opened two other boxes, she attempted to open the third box, which turned out to be
impossible, after which the father declared in a mumbling way the not-falsified prediction as
valid.
\end{footnotesize}

The assumption that the result of having 1 or 0 at the opening one of the three boxes, say A, is predetermined before opening this box, and therefore independent of wether one opens A together with B or together with C, corresponds to the assumption in classical physics. This assumption leads to the prediction that the probability to get the same result in each box is $P(x,y|x=y)=1/3$ whichever pair of boxes is opened, and whether the prophet uses two diamonds or only one.

By contrast the prediction that if one opens any of the three possible pairs of boxes one always gets the result 1 for one box and 0 for the other, would be analogous to the quantum mechanical one. In other words the probability to get the same result in each box is $P(x,y|x=y)=0$ whichever pair is opened.

It has been shown that Specker's parable offers a narrative thread that weaves together a large number of results regarding quantum contextuality, nonlocality, and complementarity \cite{Liang11}. The parable proves highly appropriate for the sake of this paper as well, in particular to clarify the relationship between contextuality and nonlocality.

This quantum ``contextual magic" in the parable can be interpreted in three main different ways: Superdeterminism, Collapse, Many-Worlds.

\section {Superdeterminism} \label{super}

The prophet knows which pair of boxes the suitor will choose, for instance B and C, and prepares the setting with a diamond in only one of these two boxes so that either result 10 or 01 appears. This interpretation amounts to deny human free will.

In the context of the discussion about ``Infuturabilien" (divine omniscience and human free will) Superdeterminism would correspond to the solution that God knows which choice we will make, together with its outcome, before we do it.

\section {Copenhagen}\label{copen}

The outcome results become determined at the moment of measurement (opening of two boxes and observing whether each of them contains a diamond or not). There is the freedom on the part of the experimenter to choose the settings of the apparatus (one of the three possible pairs of boxes) and there is the freedom on the part of nature in choosing the outcome results (either 10 or 01) for each single event i.e: each suitor's choice. The fact that each measurement realizes only one of several possible results described by the ``wave function" is usually referred to as the ``collapse of the wave function". Nonetheless this term conflates two different things:

a) The decision about which of the possible results becomes realized.

b) The fact that this results becomes ``irreversibly recorded" and can be observed.

According to the orthodox view the moment of ``detection" is crucial.

The assumption that a) (the decision of the outcome) happens at detection implies ``nonlocal" coordination between detection events. As it is well known, Einstein argued against this ``nonlocality" as early as 1927 \cite{bv}. The quantum collapse, he claimed, implies coordination of the detectors, which cannot be explained by influences propagating with velocity $v\leq c$; this involves ``an entirely peculiar mechanism of action at a distance, which [...] implies to my mind a contradiction with the postulate of relativity." \cite{bv}.

We will discuss quantum nonlocality more in detail later in Section \ref{nl-context}. Here it is suitable to remember what experiments in the last three decades have demonstrated and Specker's parable also illustrates (see Fig.\ref{f4}): Quantum correlations cannot be interpreted in the sense that one event occurs first and is the cause of the other, but one should rather state that correlated events arise like a single event coming from outside space-time; the prophet's assignment of outcomes does not involve any time order and so can be considered covariant. Actually both, quantum nonlocal and local relativistic correlations, assume ``free will" and happen without connection in space-time.\cite{as15}

The fact that at ``detection" something ``irreversible" happens (assumption b) above) has been emphasized by John A. Wheeler's: ``No elementary quantum phenomenon is a phenomenon until it is a registered (`observed', `indelibly recorded') phenomenon, `brought to a close' by `an irreversible act of amplification'." \cite{Wheeler97}. ``Irreversibility" is the core of the so called ``measurement problem" and has the noteworthy implication that we cannot apply quantum mechanical description to visible objects: The two detectors watching the output ports of a Mach-Zehnder interferometer can be considered ``nonlocally" coordinated (see Section \ref{nl-context} hereafter), but one detector cannot be considered as being in superposition of two distant locations. Niels Bohr himself postulated the ``necessity of discriminating in each experimental arrangement between those parts of the physical system considered which are to be treated as measuring instruments and those which constitute the objects under investigation" and referred to this necessity as ``\emph{a principal distinction between classical and quantum mechanical description of physical phenomena}" (italics by Bohr) \cite{nb}. Admittedly, Bohr seems to weaken this distinction by stating that: ``It is true that the place within each measuring device where this discrimination is made is in both cases [classical and quantum] largely a matter of conveience." But he insists that in quantum theory the distinction is of ``fundamental importance": The measuring device has to be classical even if the classical theories do not suffice in accounting for the new types of quantum regularities.\cite{nb} So, Bohr seems to suggest that there are definite conditions defining when a result appears and can be observed, even if for the time being we don't know which these conditions are.

This means that Erwin Schr\"{o}dinger with his famous ``cat paradox" has in fact popularized a confused interpretation of quantum theory. Defining ``standard quantum theory" as a theory ``which does not impose any constraints on the complexity of objects it is applied to", or assuming that ``an agent A can apply the rule to arbitrary systems S around her, including large ones that may contain other
agents [system S can even be the entire lab]"(as assumption (\textsf{Q}) in \cite{fr} states) seems unfitting. ``Standard quantum theory" -at least in Wheeler's understanding- imposes constraints on the complexity of objects it is applied, as soon as this complexity becomes equal as or larger than that involved in a detection: \emph{If a system collapses, it collapses for all observers.}

However \cite{fr} has the merit of showing that any theory accepting both, Born's rule and description of ``macroscopic" objects (like experimenters) in terms of quantum superposition, is not consistent. This result is relevant for our discussion later in Section \ref{cat}.

\begin{figure}[t]
\includegraphics[trim = 2mm 85mm 10mm 0mm, clip, width=0.99\columnwidth]{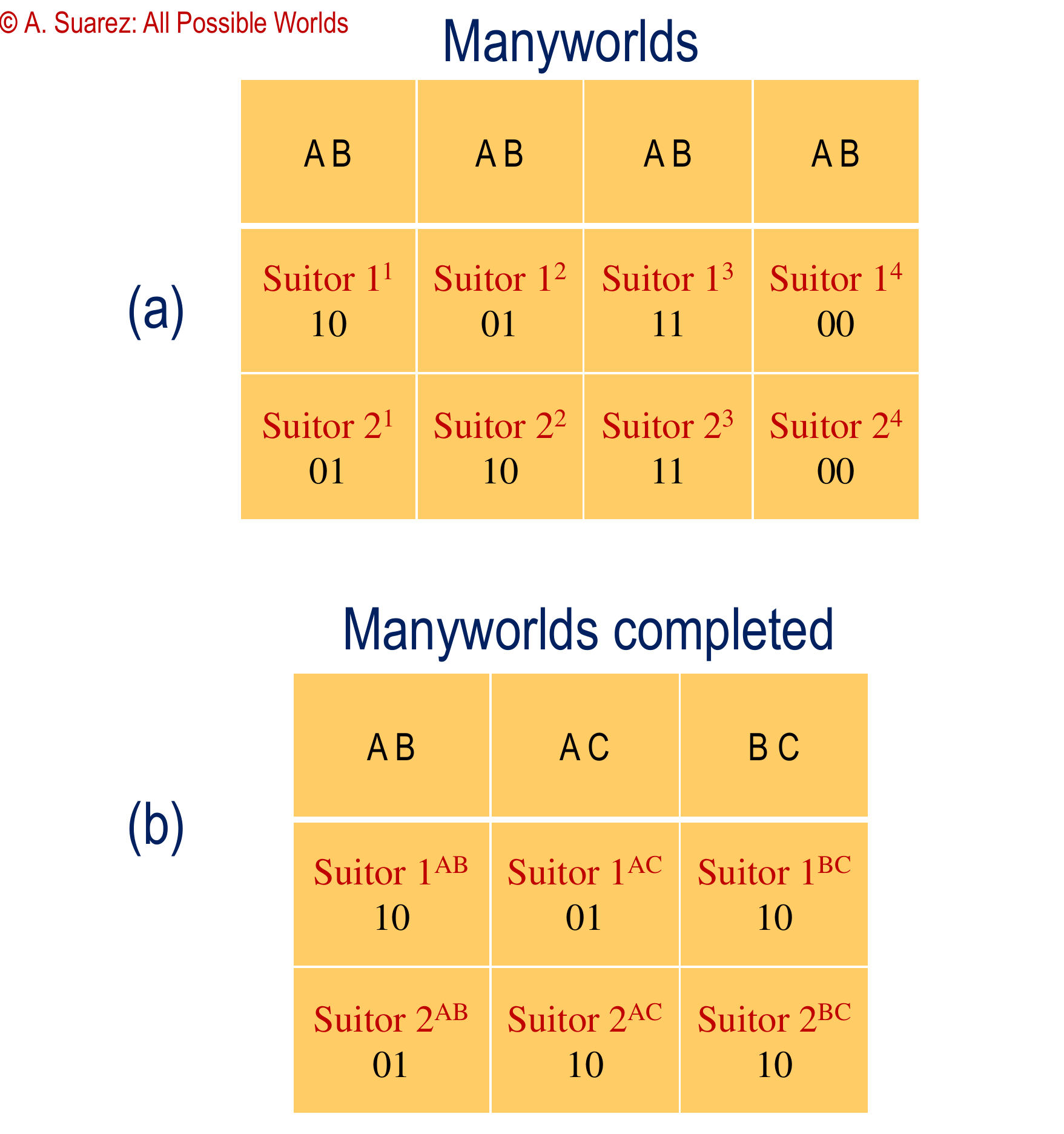}
\caption{\textbf{Many-Worlds:} (a) At the opening of the boxes AB the world splits in four worlds. In each of these one of the four possible results becomes realized and observed by a corresponding suitor. (b) Many-Worlds completed with all possible choices.}
\label{f2}
\end{figure}

\section {Many-Worlds}\label{mw}

This interpretation (represented in Fig.\ref{f2}a) goes back to the relative state formulation of Quantum Mechanics proposed by Hugh Everett 1957 \cite{he}. ``The fundamental idea of the MWI [...] is that there are myriads of worlds in the Universe in addition to the world we are aware of. In particular, every time a quantum experiment with different possible outcomes is performed, all outcomes are obtained, each in a different world, even if we are only aware of the world with the outcome we have seen."\cite{lev}.

At any choice the world splits in 4 parallel worlds, so that all possible outcomes become realized although in different parallel worlds and is seen in each world by a corresponding observer (``suitor"), as represented in Fig. \ref{f2}(a) for choice AB. The parallel worlds and suitors resulting at each experiment are in principle ``experimentally" inaccessible to each other.

Many-Worlds has been formulated in various ways. David Deutsch uses the ``Multiverse" formulation to explain the notion of a ``quantum computer" \cite{deutsch}, which contains only possible outcomes predicted by quantum mechanics: In the case of Fig. \ref{f2}(a) the ``Multiverse" consists in only two universes represented by the first two colones.

The meaning of Many-Worlds is particularly well brought to light in the formulation ``Parallel lives" by Gilles Brassard and Paul Raymond-Robichaud \cite{Brassard}. According to this version: ``When Alice pushes a button on her box, she splits in two, together with her box. One Alice A sees the red light flash on her box, whereas the other A* sees the green light flash. Both Alices, A and A*, are equally real. However, they are now living parallel lives: they will never be able to see each other or interact with each other. In fact, neither Alice is aware of the existence of the other, unless they infer it by pure thought as the only reasonable explanation for what they will experience when they test their boxes." \cite{BRR1}

Variants and extensions of Many-Worlds are properly characterized by Frauchiger-Renner's statement: ``Their common feature is that they do not postulate a physical mechanism that singles out one particular measurement outcome, although observers have the perception of single outcomes." (\cite{fr}, v1) This characterization points also to what may be a main inconsistency of Many-Worlds:

Indeed, according to Leibniz's principle, ``if there is no possible perceptible difference between two objects, then these objects are the same, not superficially, but fundamentally" \cite{BRR2}. As far as one keeps to this principle, if Alice in our world can never be able to see the other Alice*, then one should conclude that Alice* has no physical reality at all: Things that cannot in principle be perceived by the senses do not exist \emph{within space-time}. And if the existence of Alice* can be inferred by reasoning but cannot in principle be perceived by the senses, this means that Alice* exists outside space-time.

But one could object: Wouldn't the latter statement imply that events that happened in the distant past, about which we just read in history books, but which we cannot (and could not) perceive with our senses, do not exist?

We access such events through observations we perform today: archaeological vestiges or writings documenting them (see British museum).
So these events happened in our space-time. A past event that is in principle inaccessible to observations we can perform today does not belong to the physical reality we can describe.

The Alice* in the ``parallel-lives" interpretation is something we cannot access through any observation in our world but only through reasoning. Accordingly, Alice* lives outside our space-time, and invoking Specker's parable she can be considered existing in ``God's mind", as I propose in the following Section \ref{apw} with All-Possible-Worlds.

In summary, ``Copenhagen" and ``parallel lives" are basically equivalent: Both interpretations imply that the ``physical reality" we live in is more than what we can access with our senses.

\begin{figure}[t]
\includegraphics[trim = 2mm 115mm 10mm 0mm, clip, width=0.99\columnwidth]{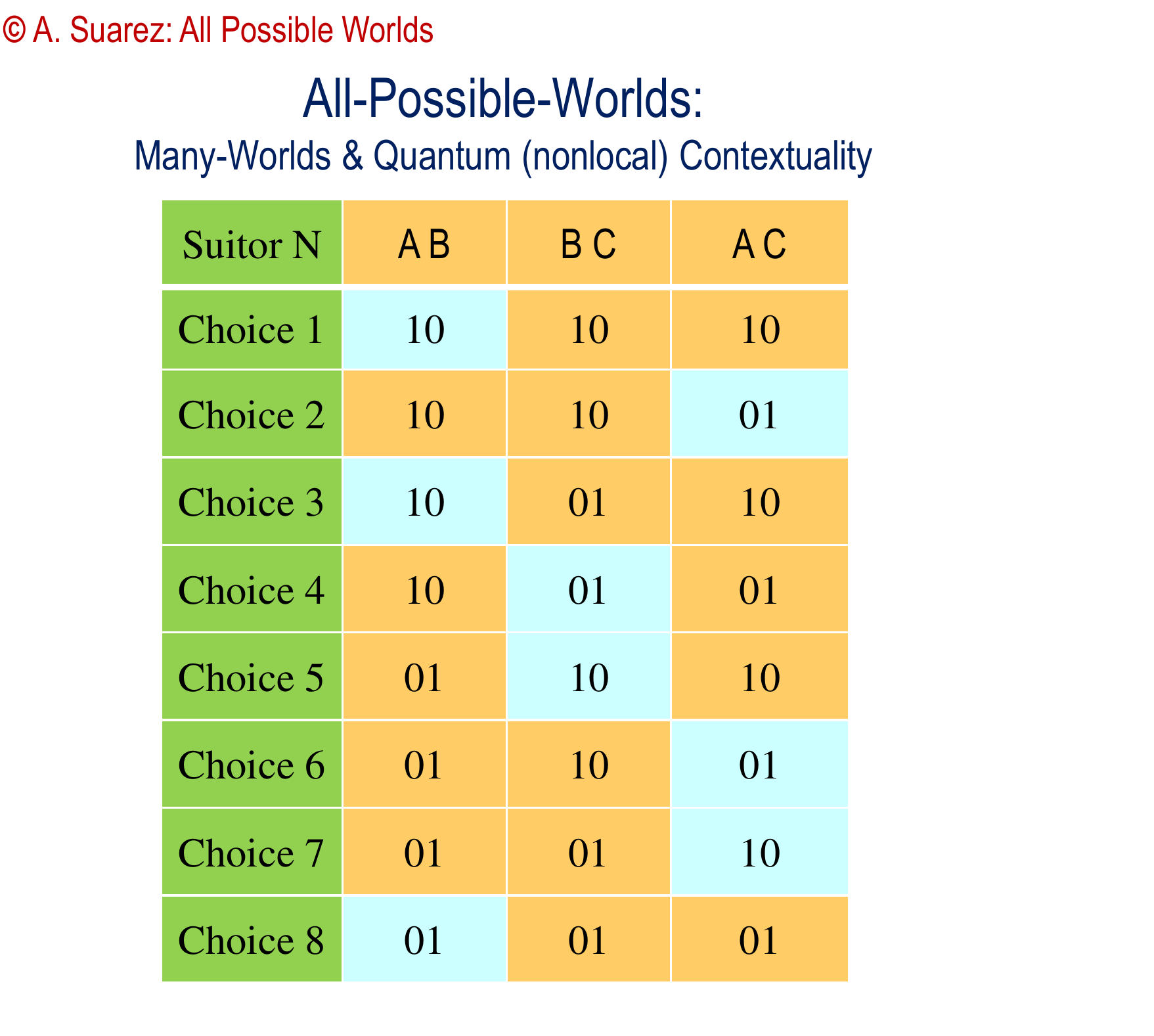}
\caption{\textbf{All-Possible-Worlds:} The ``prophet" has in his mind an outcome choice for each of the three possible choices a candidate can do. Each suitor is free to make his choice as he wants. The prophet let then appear either the result 10 or 01. The sky-blue boxes represent the history of choices and outcomes of Suitor N, who was allowed to make 8 choices at different times. Other suitors would have different histories. For any choice the prophet's assignment is nonlocal, as illustrated in Fig.\ref{f4}.}
\label{f3}
\end{figure}

\section {All-Possible-Worlds}\label{apw}
Many-Worlds to be consistent should also consider the possible choices the experimenter can do. In the context of the parable this would mean that if the suitor makes choice AB, parallel worlds arise where ``clones" of the suitor make choice BC, respectively AC, as represented in Fig.\ref{f2}(b). Paraphrasing Nicolas Gisin one could say that in Many-Worlds the experimenter should not be merely a passive observer, but play an active role \cite{ng17}.

Accordingly each single choice of a box pair would lead to 12 parallel worlds.

But then one could as well think of a ``razored" version keeping only the parallel worlds corresponding to the different choices experimenters can do as represented in Fig.\ref{f2}(b), and renounce to the splitting for performing all possible outcomes as represented in Fig.\ref{f2}(a).

This completed version of Many-Worlds leads straightforwardly to our proposal of ``All-Possible-Worlds":
In the context of Specker's parable the ``prophet" has in his mind a well-defined outcome choice for each of the three possible choices a particular suitor can make, as represented in Fig.\ref{f3}. And for many rounds of a same experiment (say AB) the outcomes (either 10 or 01) the suitor would observe are distributed according to the quantum mechanical predictions (``the Born rule"). Suppose each candidate is allowed to make different choices at different times: In Fig.\ref{f3} is represented the history of choices and outcomes corresponding to Suitor N. Nonetheless, one can think that the results the prophet assigns to these choices may depend on other circumstances: so for instance if instead of Suitor N a different Suitor N' had made the choices, the results N' would have obtained are not necessarily the same as those Suitor N obtains, although they may have been also distributed according to Born rule. In this sense we do not postulate \emph{a physical mechanism} that \emph{singles out} one particular measurement outcome. However once measurement ``irreversibly" happens, then the outcome becomes registered and can be accessed by any observer (however ``irreversibly" may have ``exceptions", which are discussed more in detail later in Sections \ref{second} and \ref{cat}).

According to ``All-Possible-Worlds" the decision of the outcome (the ``collapse", understood as a) in Section \ref{copen} previously) occurs in the prophet's mind (when he assigns an outcome to a determined measurement), and the opening of a box-pair (the apparatus) reveals the result the prophet assigns. Nonetheless, since we cannot in principle access the prophet's mind the decision can be said to occur at detection ``for all practical purposes", but it cannot properly be said that detection ``singles out" the outcome.

``All-Possible-Worlds" illustrates well what quantum contextuality means: The prophet assigns results to each possible choice and each possible round, however the assignment is not done for each single box but \emph{to each possible pair of boxes jointly}. Consequently for each round the assignment for A depends on whether A is ``measured" together with B (experiment AB), or with C (experiment AC).

This interpretation has a remarkable number of implications we discuss in the following.

\section {Quantum nonlocality and contextuality}\label{nl-context}

Bell inequalities establish a limit to the degree of correlation originating from a cause in the common past cone of the correlated events. Consequently correlations between space-like separated events that violate Bell inequalities cannot be explained by local-relativistic influences propagating at velocity $v\leq c$. This fact is referred to as \emph{Bell nonlocality} and is characteristic for entanglement experiments.

\begin{figure}[t]
\includegraphics[trim = 2mm 110mm 2mm 22mm, clip, width=0.99\columnwidth]{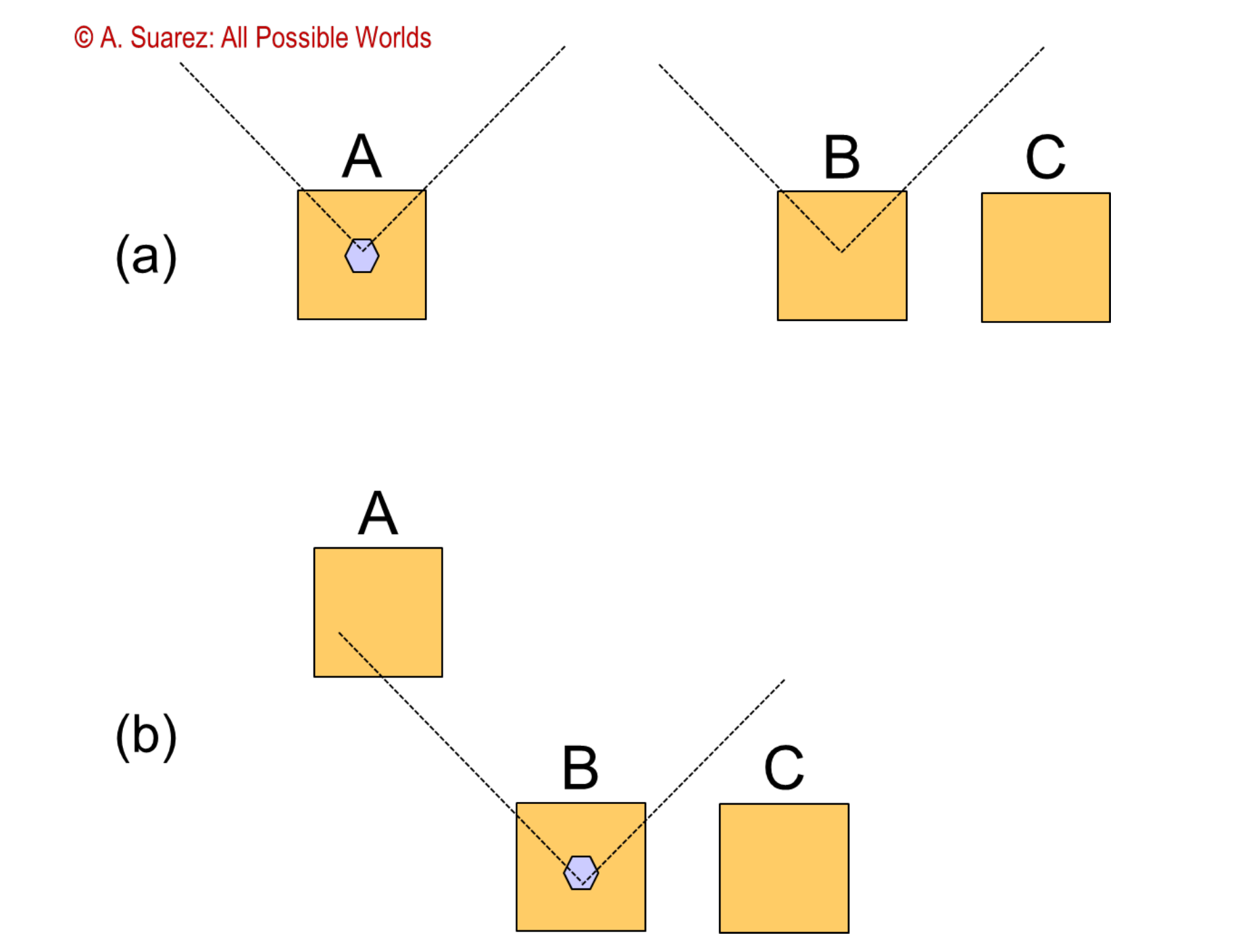}
\caption{\textbf{Contextuality and Nonlocality:} (a) the opening of the boxes are space-like separated events. (b) The opening of the boxes are time-like separated events: Contextual correlations can be simulated by a classical machine sending a signal from B to A, propagating in space-time at $v\leq c$.}
\label{f4}
\end{figure}

This also means that in such experiments one cannot assign a result for Alice's measurement independently of Bob's measurement. In this sense it holds that:

\begin{eqnarray}
\textsf{\emph{Bell nonlocality}} \implies \textsf{\emph{Contextuality}}
\label{1}
\end{eqnarray}

On the other hand contextuality appears in experiments with \emph{single} particles of $spin \geq1$, where no Bell inequalities can be defined, and therefore it holds also that one can have contextuality without Bell nonlocality:

\begin{eqnarray}
\textsf{\emph{Contextuality}} \centernot \implies \textsf{\emph{Bell nonlocality}}
\label{2}
\end{eqnarray}

Conversely, \emph{Bell locality} means fulfillment of Bell inequalities and therefore is a feature applying to entanglement experiments with 2 or more particles corresponding to Hilbert spaces of $d \geq 4$, \emph{d} non prime, while \emph{locality} refers to the fact that correlations can be explained by influences propagating at velocity $v\leq c$ and may happen in any quantum experiment.

Since non-contextuality means that the measurements Alice performs are independent of those Bob performs it is obvious that:

\begin{eqnarray}
\textsf{\emph{Non-contextuality}} \implies \textsf{\emph{Locality}}
\label{3}
\end{eqnarray}

Assumption of non-contextuality leads to non-contextual inequalities, and accordingly, contextuality is supposed to be demonstrated by results violating such inequalities in single qutrit experiments.\cite{guhne, alonso, ac17}

If one restricts the meaning of ``nonlocality" to ``violation of Bell inequalities" in entanglement experiments with two or more particles (systems corresponding to Hilbert space of $d \geq 4$, \emph{d} non prime), then one is tempted to conclude also that quantum contextuality may be a better feature than ``nonlocality" for distinguishing quantum from classical, because contextuality applies to all quantum experiments excepted single qubits (systems corresponding to Hilbert space of $d \geq 3$).

However Specker's parable has the merit to show that such a generalization is misleading. In fact the boxes can be placed at a distance from each other so that the openings of the boxes are time-like separated events as sketched in Fig.\ref{f4}(b), then contextual correlations could be simulated by a \emph{classical} mechanism, that is, by causal relativistic connections between the boxes propagating with $v\leq c$) or information contained within space-time. Thus contextual results (i.e.: experimental violation of non-contextual inequalities) for them alone cannot rule out \emph{all} classical features, but only if they appear when the openings are space-like separated as sketched in Fig.\ref{f4}(a). In other words in single qutrit experiments it holds that:

\begin{eqnarray}
&& \textsf{\emph{Violation of non-contextual inequalities}} \\ \nonumber
&& \;\; \wedge \;\; \textsf{\emph{Space-like separated detection events}}\\ \nonumber
&&\implies \textsf{\emph{Non-classicality}}=\textsf{\emph{Quantum}}\;\;\;\;\;\;\;\;\;\;\;
\label{4}
\end{eqnarray}

\ \\
very much the same way as in entanglement experiments with 2 qubits it holds that:

\begin{eqnarray}
&& \textsf{\emph{Violation of Bell inequalities}}\\ \nonumber
&& \;\; \wedge \;\; \textsf{\emph{Space-like separated detection events}}\\ \nonumber
&&\implies \textsf{\emph{Non-classicality}}=\textsf{\emph{Quantum}}\;\;\;\;\;\;\;\;\;\;\;
\label{5}
\end{eqnarray}

\subsection {Nonlocality at detection}\label{nl-detection}

To understand the relationship between contextuality and nonlocality one should keep in mind (see Section \ref{copen}) that according to standard quantum mechanics the decision of the outcome happens at the moment of detection (``wavefunction collapse'').

The reason for this assumption are interference experiments with single particles as for instance those using a Mach-Zehnder interferometer with two output ports monitored by corresponding detectors D0 and D1: If the outcome (i.e.: which of the two detectors counts) were determined before detection by the path the particle travels, then half of the time D0 should count and half of the time D1, and the interference pattern would disappear.

As referred to in Section \ref{copen}, at the 5$^{th}$ Solvay conference (1927) Einstein objected to the assumption of ``decision at detection" by means of a \emph{single-particle} gedanken-experiment \cite{bv}. Astonishingly Einstein's gedanken-experiment in 1927 has been first realized using today's techniques in 2012 \cite{Guerreiro12}. In this experiment single photons impinge into a beam-spitter BS and thereafter get detected. The two detectors A and B monitoring the output ports of BS are located so that the decision ``to count" or ``not to count" at A is space-like separated and therefore \emph{locally} independent from the decision ``to count" or ``not to count" at B.

The experiment tests and rules out the assumption that this correlation can be explained by some sort of local coordination through signals with $v\leq c$: Even if the measurement of $P_1$ is \emph{locally independent} of the measurement of $P_2$, both measurements yield correlated results. On the other hand this nonlocal coordination cannot be used to signal faster than light from one detector to the other. The experiment also highlights something Einstein did not mention: Nonlocality is necessary to preserve such a fundamental principle as \emph{energy conservation} \cite{Guerreiro12}.

Using the standard formalism the experiment in \cite{Guerreiro12} is described as follows: The two output ports define a basis of orthogonal unitary vectors $|v_1\rangle$ and $|v_2\rangle$ in a Hilbert space with $d=2$. A count in detector A (respectively no count in A) means a measurement that projects the quantum state into the unitary vector $|v_1\rangle$ yielding result 1 (respectively result 0), and similarly for detector B and unitary vector $|v_2\rangle$). The two measurements are associated to the projection operators: $P_1=|v_1\rangle\langle v_1|$ and $P_2=|v_2\rangle\langle v_2|$, each of them with eigenvalues 1 and 0. Since $P_1$ and $P_2$ \emph{commute} both can have definite values, and the measurements of these operators are said to be \emph{compatible}. Additionally, if the two orthogonal projectors are measured on the same system they yield correlated results: one of the two measurements must give result 1 and the other result 0. This can be considered the simplest form of ``exclusivity" as referred to in \cite{ac08}.

This result has also the following deeper meaning: The two detectors A and B should be considered as a single measuring apparatus and the \emph{jointly} result of the corresponding detection events should be considered as a single measurement result even when these events are space-like separated. The result is not for instance: ``detector A counts", but rather: ``detector A counts \emph{and} detector B doesn't count".

The 5th Solvay Conference was attended also by Louis de Broglie. He presented an interpretation different from the Copenhagen one, the so called ``pilot-wave" or ``empty-wave" picture: The particle always follows a well determined path from the source to a detector, but there is an \emph{undetectable} ``pilot-wave" that travels by the alternative paths, joins the particle at the arrival and, taking account of the path-length difference, guides the particle to one or other of the detectors to producing the characteristic interference fringes predicted by Quantum Mechanics.

Unfortunately, in the context of the single-particle gedanken-experiment Einstein presented at the Conference, de Broglie’s ``empty wave" can easily be misunderstood as sort of (energyless) “ghost-wave” carrying only information but travelling at the same speed of the material particle, so that you can explain interference avoiding nonlocal coordination. Indeed, Einstein misinterpreted this way de Broglie’s idea and praised it as going in ``the right direction".\cite{bv}

The debate resumed in 1935 with the famous EPR paper, where Einstein proposed a new thought experiment with two particles, that is, an entanglement experiment. Some years later (1952) David Bohm published an article \cite{db} applying de Broglie’s picture to explain the EPR experiment. In this context it became clear that the ``empty-wave" has nothing to do with a ``ghost-wave" propagating locally within the ordinary 3-space, but is rather a mathematical entity defined in the so called ``3N-space or configuration space"(\cite{jb} p. 128). In other words, the ``wave'' guides the material particle from outside the ordinary space-time (nonlocally) to produce space-like separated correlated detections at the two sides of the setup. This is the origin of the Bohmian picture, which inspired John Bell to find a criterion allowing to decide between quantum non-locality and Einstein’s locality by means of entanglement experiments: The meanwhile famous Bell inequalities \cite{jb}. Since 1982 a number of well-known experiments have ruled out Einstein’s locality upholding the quantum mechanical predictions.

The important lesson of the story is that the ``de Broglie-Bohm" interpretation is itself nonlocal, although it can be misunderstood as local in single-particle experiments. The only difference with relation to Copenhagen is, that in this picture the ``wave-function" influences macroscopic material objects like detectors, and can be interpreted as inducing nonlocal coordination between detection events, and in the ``de Broglie-Bohm" picture the ``wave-function" influences microscopic material particles, and can be interpreted as inducing nonlocal coordination between choices at the beam-splitters.

The conclusions from the space-like single qubit experiment can be straightforwardly extended to a qutrit experiment using particles of spin 1 crossing a Stern-Gerlach non-homogeneous magnetic field (i.e: experiments like those invoked in the \emph{Kochen\&Specker theorem}\cite{ks67}). Suppose each of the three output ports is monitored by a corresponding detector D0, D1 and D2. We denote the corresponding three basis states as $|v_0\rangle$, $|v_1\rangle$, $|v_2\rangle$. The three projectors: $P_0=|v_0\rangle\langle v_0|$, $P_1=|v_1\rangle\langle v_1|$, $P_2=|v_2\rangle\langle v_2|$ are orthogonal and commute pairwise, and therefore also according to QM they can have jointly sharp defined values. Additionally, the three orthogonal projectors $P_0$, $P_1$, $P_2$ yield correlated results: One of the three measurements must give result 1 (count) and the other two result 0 (no-count); this property is referred to as ``exclusivity principle" in contextuality \cite{ac08}. Suppose now the three detectors are set in such a way that the detection events are pairwise space-like separated. Analogously to the qubit, the correlations (``exclusivity") mean nothing other than \emph{conservation of energy} ensured by nonlocal coordination of detections. And also here the three detection events should be considered a single measurement result as for instance: ``detector D0 counts \emph{and} detectors D1 and D2 don't count". If one changes the basis (by rotating the Stern-Gerlach) the detectors corresponding to the new basis define a completely different apparatus, even if the two basis have a projector in common.

\begin{figure}[t]
\includegraphics[trim = 2mm 185mm 2mm 0mm, clip, width=0.99\columnwidth]{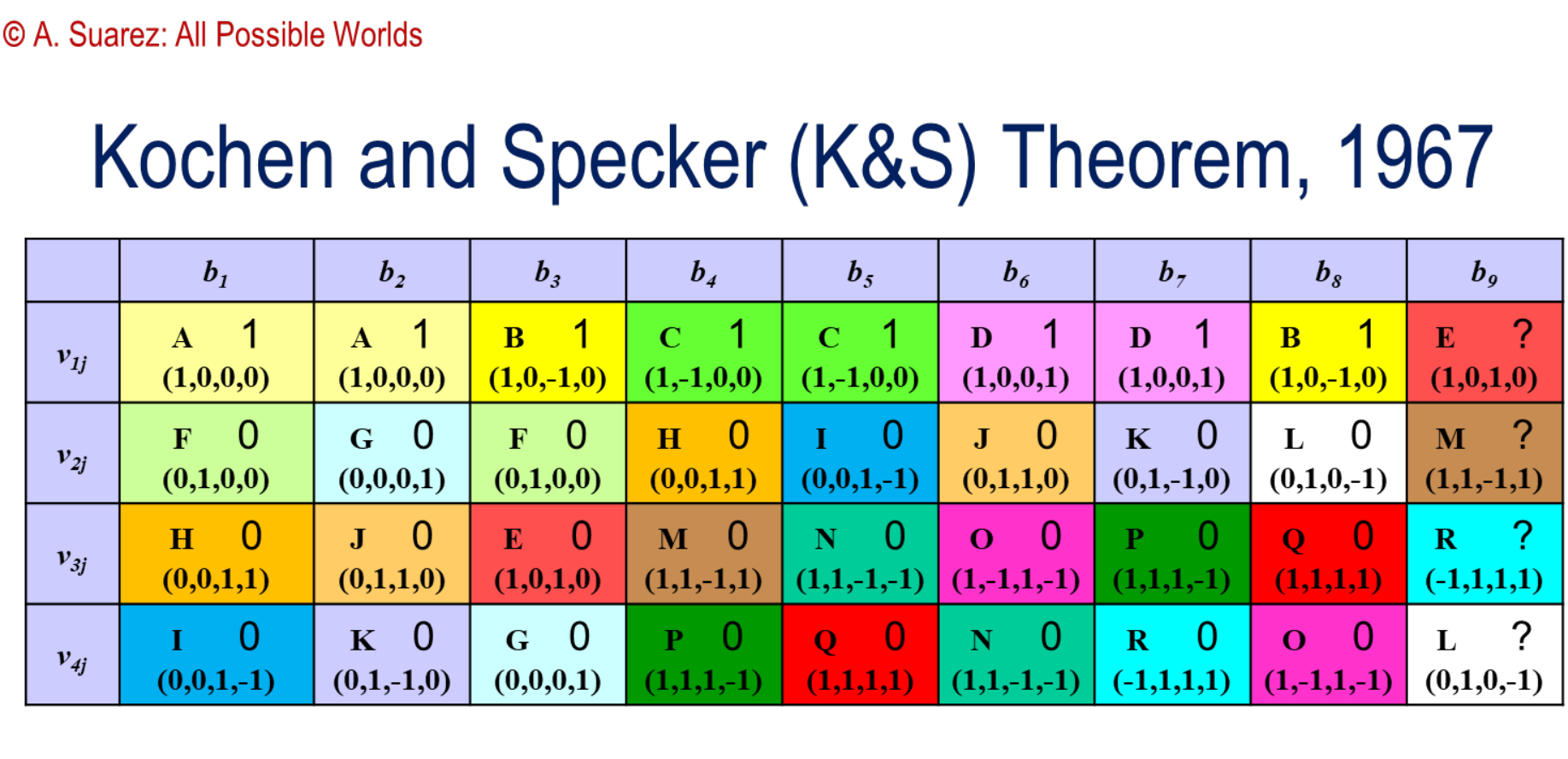}
\caption{\textbf{Proof of K\&S with 18 rays, according to \cite{ac96}:} One arranges 18 unitary vectors in 9 different bases of 4 pairwise orthogonal unitary vectors $v_{i,j},i\;\in\{1,4\}, j\;\in\{1,9\}$, so that each vector appears in two different basis (boxes denoted by same letter and color). Projection operators $P_{ij}=|v_{ij}\rangle\langle v_{ij}|$ onto each of the 18 unitary vectors yield either value 1 or 0, indicated in the top right corner of each box. Since each $P_{ij}$ appears repeated in two different bases one has: $P_{19}=P_{33}, P_{29}=P_{34}, P_{39}=P_{47},P_{49}=P_{28}$. This means: either each $P_{i9}=0$ and then $\sum_{i=1}^4 P_{i9}=0$, or one of the four $P_{i9}$ yields value 1, say $P_{39}=1$, and then $P_{47}=1$ with $\sum_{i=1}^4 P_{i7}=2$. In any case one is led to a contradiction with (\ref{Pij})}
\label{f5}
\end{figure}

More in general, the \emph{standard} postulate that ``observables corresponding to a complete set of orthogonal projection operators have \emph{simultaneously} definite measurement values" does not mean that the value can be assigned to each single detector separately of the others. Each projector is not an entity for it alone, separate from the other commuting projectors defining the basis of the experiment (this is also the message quantum contextuality conveys, as we will see in the next Subsection \ref{context-class}). The ``weirdness of the quantum" originates from the `` classical" prejudice that one can define ``variables" which are independent from each other.

Notice that in the preceding reasoning it is crucial to assume that the measurements at D0, D1 and D2 are pairwise space-like separated. Otherwise the correlations could be explained \emph{classically}, that is by means of relativistic local-causality or influences propagating at velocity $v \leq c$. Certainly, one could explain the qutrit experiment also locally by means of Einstein's ``ghost-waves", as those referred to previously, but the explanation would fail for experiments with 2 entangled qutrits.

In experiments with 2 or more particles, nonlocality at detection appears in the form of Bell nonlocality, that is, as correlations between space-like separated events that violate Bell inequalities; Bell nonlocality ensures conservation of linear and angular momentum.

Notwithstanding that nonlocality came to fame with the discovery of Bell's theorem \cite{jb}, it would be awkward that such a crucial feature prevailed only in multipartite systems and hence failed in systems with a prime number of states (Hilbert spaces with prime dimension), played a decisive role in 2 qubits systems corresponding to Hilbert spaces with $d=4$ but had no significance in single particle spin $3/2$ systems, which also correspond to Hilbert spaces with $d=4$. Such oddities disappear as soon as one acknowledges (in line with the ``collapse postulate") that ``nonlocality" rules the \emph{whole} quantum realm, from the single qubit or two-state systems (Hilbert space with $d=2$) onwards, in the form of ``nonlocality at detection". In favor of this view also speaks the fact that experimental demonstration of Bell nonlocality requires space-like separated detection events. ``Nonlocality at detection" \cite{Guerreiro12} is the fundamental and general form of ``nonlocality", and historically also the first one.

\subsection {Quantum contextuality and ``classicality"}\label{context-class}

The assumption that each projector is an entity for it alone, separate from the other commuting projectors defining the basis of the experiment bears a contradiction. This is the theorem Simon Kochen and Ernst Specker proved (1967) in Hilbert space of $d=3$
using 117 rays (projectors) \cite{ks67}. Later simplified versions of the theorem have been produced. In 1991 a proof was provided with 33 rays also for $d=3$. The minimal possible K\&S set consists of 18 rays in Hilbert space of $d=4$ \cite{ac96, ac08}, and leads to the particularly simple proof reproduced in Fig. \ref{f5}:

One defines 18 projection operators $P_{ij}=|v_{ij}\rangle\langle v_{ij}|$ onto each of the 18 unitary vectors. Because of the orthogonality relations it holds that for each basis the projection operators commute pairwise, and when measured, one of them must yield value 1 and the other three value 0:

\begin{footnotesize}
\begin{eqnarray}
[P_{ij},P_{kj}]=0; \; i,k \in\{1,4\}, \;i \neq k; j\;\in\{1,9\}
\label{Com}
\\
P_{ij}= a,\; a\in\{0,1\}, \;\;\;\;\sum_{i=1}^4 P_{ij}=1
\label{Pij}
\end{eqnarray}
\end{footnotesize}

Consider now a non-contextual theory that assigns a value to each $P_{ij}$ \emph{independently} of the basis in which it is measured. Since each $P_{ij}$ appears repeated in two different basis in Fig. \ref{f5} one is led to a contradiction with (\ref{Pij}).

To test experimentally this contradiction one introduces observables $A_{ij}=I-2P_{ij}= a, \; a\in\{-1,1\}$, which by definition fulfill the compatibility condition: $[A_{ij},A_{kj}], \; i,k \in\{1,4\}, \;i \neq k$. Non-contextual theories assuming that measurement of $A_{ij}$ is not correlated with other ``compatible measurements" fulfill the following inequality \cite{ac08}:

\begin{footnotesize}
\begin{eqnarray}
\sum_{j=1}^9 A_{1j}\cdot A_{2j}\cdot A_{3j}\cdot A_{4j} \leq 7
\label{NCI}
\end{eqnarray}
\end{footnotesize}

This inequality is violated by quantum theory, which predicts:

\begin{footnotesize}
\begin{eqnarray}
\sum_{j=1}^9 A_{1j}\cdot A_{2j}\cdot A_{3j}\cdot A_{4j} = 9
\label{QI}
\end{eqnarray}
\end{footnotesize}

In the context of the Stern-Gerlach experiment with single spin 1 particles the theorem has the following meaning: Theories assuming values that are a) \emph{predetermined} before measurement, and b) \emph{non-contextually} assigned, cannot reproduce the correlations appearing in this single qutrit experiment.

Indeed non-contextual assignment of values implies that for some bases two of the three detectors would simultaneously count. This leads to a non-contextuality inequality similar to that in (\ref{NCI}). Violation of this inequality proves that the values are not assigned to each single projector for it alone but takes account of the basis of measurement.

Nevertheless this contextuality could in principle be explained \emph{classically} by means of \emph{local} influences between detection events \emph{as far as these events are time-like separated}.

This shows that an ideal contextuality experiment expands the conditions of the qubit experiment in \cite{Guerreiro12} to a qutrit one: In fact in both experiments one tests whether the number of coincidence counts is significant larger than noise and fortuitous 2-particle events when detections happen space-like separated. And in both experiments lack of coincidences prove that conservation of energy in single quantum events is granted by ``nonlocal coordination" of detections. And this seems to me is the ``physical justification" of contextuality Ernst Specker and Simon Kochen were longing for \cite{ac09}.

Regarding inequality (\ref{NCI}) it has been claimed that``[Its] derivation is similar to a standard derivation of a Bell inequality. The only difference is that in a Bell inequality we assume that the result of a measurement of $A_{ij}$ is independent of spacelike separated measurements, while here we assume that it is independent of compatible measurements." \cite{ac08,guhne}

Nonetheless, in the light of our preceding analysis one should correctly rather state: In both cases, Bell and non-contextuality inequalities, the derivation rests on assignment of values to each projector for it alone, without taking account of the basis of measurement. And in both cases, Bell and contextuality experiments, space-like separated detection events are necessary to rule out any \emph{classical} explanation of observed correlations by means of causal relativistic connections (signals propagating with $v \leq c$), that is, information contained within space-time.

One may be tempted to object: ``The correlations are exactly the same regardless whether there are or not space-like separation. Quantum mechanics is still there even if you are poor to achieve space-like separation." But, how can you know if you don't perform the ``space-like separation" experiment?: ``Unperformed experiments have no results". So one can't help calling on \cite{Guerreiro12} for help and extrapolate this result to argue that any experiment using space-like separated detectors can be supposed to yield the same correlated results as experiments with time-like separation.

One should also be careful in using the concept of ``independence". Undoubtedly, if Alice`s measurement at A and Bob`s measurement at B are uncorrelated (randomly distributed) like the tossing of two fair coins, then we can deduce that they are ``independent" from each other, or ``happen without disturbing each other": neither event can be considered to cause the other one.

However if both measurements prove to be correlated \emph{beyond the bound ``biased coins" or ``non-contextually assigned values" } can produce, what can we deduce?:

If the distance A-B is such that the tossing at A happens time-like before the tossing at B (as it is the case in experiments with sequential measurements), then we cannot exclude that the correlations are produced by some signal traveling with velocity $v \leq c$ from A to B.

By contrast, if the distance A-B is such that the tossing at A is space-like separated from the tossing at B we can conclude that the correlation cannot be explained by any connection in space-time, even if one can say that neither Alice's result depends on Bob's one, nor Bob's result depends on Alice's one: The correlated outcomes form a joint result coming from outside space-time.

In this sense the concept of ``independence" alone seems not helpful to distinguishing \emph{quantum contextual} correlations from \emph{classical} ones, and consequently  experiments with sequential measurements don't rule out \emph{classical} explanation for demonstrated contextual correlations.

This applies in particular to the ongoing test presented in \cite{alonso}. The experiment uses a single trapped-ion qutrit to demonstrate the violation of a non-contextuality inequality which is the qutrit counterpart of that in (\ref{NCI}). The qutrit basis states are represented by three fine-structure levels in a ion ${}^{40}\texttt{Ca}^+$: $|0\rangle=|S_{1/2}(m_j = -1/2)\rangle$ in the ground-state manifold, and $|1\rangle = |D_{5/2}(m_j = -3/2)\rangle$ and $|2\rangle =|D_{5/2}(m_j = -1/2)\rangle$ in the metastable $D_{5/2}$ manifold. For the sake of our analysis it suffices to say that the measurement happens by means of only one detector monitoring the state $|0\rangle$. If this detector counts (photons scattered by the ion) the qutrit is projected onto $|0\rangle$ (which accordingly is called ``bright state"); if the detector doesn't count the qutrit is projected onto a superposition of the states $|1\rangle$ and $|2\rangle$ in the $D_{5/2}$ manifold (which accordingly are called ``dark states").

The experiment could be compared to an Stern-Gerlach experiment using a single particle spin 1 qutrit, where one monitors only the output port corresponding to the basis state $|0\rangle$: If the detector counts the qutrit is projected onto $|0\rangle$, if it does'nt count one \emph{deduces} that the qutrit is projected onto a superposition of the states $|+1\rangle$ and $|-1\rangle$. This \emph{deduction} is equivalent to assume that each of the three output ports corresponding to the three basis states is monitored by a detector.

Currently the experiment in \cite{alonso} uses sequential measurements and not space-like separated ones. Thus it does not discard in principle the possibility that one outcome determines \emph{locally} the next one and a \emph{classical} machine simulates quantum contextuality. However such a simulation requires memory: ``while simulating quantum nonlocality with classical systems requires superluminal communication [...], simulating quantum contextuality requires memory [...] or, more precisely, the ability of storing and recovering certain amount of classical information." \cite{ac17}. The experiment establishes lower limits to the number of bits of classical memory needed for that task. To increase the difficulty of simulation and ``prevent a hypothetically conspiring ion from knowing what the context of a measurement will be" one \emph{randomizes} the sequence of measured observables using a QRNG \cite{alonso}: ``a single quantum system is submitted to an unlimited sequence of measurements, \emph{randomly} [emphasis added] chosen from a finite set of measurements" \cite{ac17}. But acknowledging ``the randomness of the QRNG" means nothing other than acknowledging the result of \cite{Guerreiro12}, i.e.: nonlocality at detection.

In summary, one can hardly deny that ``relativistic-local causality" is a \emph{classical} feature. Therefore, deciding between ``\emph{quantum} correlations" and ``classical ones" requires \emph{space-like} separated (i.e: locally independent) detection outcomes.

Accordingly the claim that ``quantum contextuality [is] a natural generalization of quantum nonlocality to the case of nonspacelike separated systems"\cite{ac17} is equivocal: Contextuality is certainly more general than Bell nonlocality but less general than ``nonlocality at detection"; and to be ``quantum" contextuality has to hold for space-like separated systems.

One should avoid by all means reducing nonlocality to ``Bell nonlocality" or conflating ``classicality" and ``non-contextuality". One can claim that ``Bell nonlocality is a white dog and contextuality is the dog", nonetheless one should not forget to add that ``nonlocality at detection makes the dog", so that Implication (\ref{4}) becomes:
\vspace{-0.2cm}

\begin{eqnarray}
&&\textsf{\emph{Violation of non-contextual inequalities}}\\ \nonumber
&& \;\; \wedge \;\; \textsf{\emph{Space-like separated detection events}}\\ \nonumber
&&\implies \textsf{\emph{Nonlocality at detection}}\implies \textsf{\emph{Quantum}}\;\;\;\;\;\;\;\;
\label{6}
\end{eqnarray}

It is also worth stressing that contextual correlations demonstrated by experiment with space-like separated detections imply that ``values predetermined before measurement" are not information stored within space-time in ``classical memories", but exist nonlocally, outside space-time (in the ``prophet's mind"). In this sense quantum contextuality, non-realism, and nonlocality at detection are one and the same feature. ``Non-realism" is certainly ``a fundamental key of the world" \cite{ac16}, but it is inseparable from ``nonlocality at detection".

\subsection {Future contingents, the Classical,\\ and the Quantum}\label{conting}

The conclusions in the preceding Subsection \ref{context-class} can be strengthened on the basis of Simon Kochen's reconstruction of quantum mechanics \cite{sk15, sk17}.

The corner stone of Kochen's reconstruction is found in the following quotation referring to EPR experiments:

``How can correlations between spin components of two particles subsist when these spin components do not have values?
To understand how this can happen it is necessary to distinguish between events that have
already happened and future contingent events. Thus, for instance, if $a \vee b$  is currently true, then
either \emph{a} is true or \emph{b} is true. When future events are considered, this no longer the case: if $a \vee b$
is certain to happen, it is not the case that \emph{a} is certain to happen or \emph{b} is certain to take place."\cite{sk17}

As a famous example of this logical feature Kochen refers to Aristotle's sea battle: ``A sea-fight
must either take place to-morrow or not, but it is not necessary that it should take place to-morrow,
neither is it necessary that it should not take place, yet it is necessary that it either should or
should not take place tomorrow. Since propositions correspond with facts, it is evident that when in
future events there is a real alternative, and a potentiality in contrary directions, the corresponding
affirmation and denial have the same character." \cite{sk17}

Here one implicitly introduces the following

\emph{Definition:}
An event \emph{a} is called future contingent with relation to a future time T, if to establish whether event \emph{a} takes place or not one has to await time T.

It is interesting to compare Aristotle's example of the ``sea battle" with the statement: ``The Sun will raise tomorrow at time T at point P of the horizon".

The ``sea battle" is a future contingent event with relation to ``tomorrow".
By contrast, according to classical physics the raising of the Sun is not a future contingent event because it is assumed we can know today about it with certainty.

According to the \emph{Definition} above it is impossible even in principle to predict with certainty whether event \emph{a} will take place or not on the basis of the currently observable data, that is on the basis of information stored within space-time.

This is the same as assuming that on the basis of \emph{classical} physics we can establish neither that ``\emph{a} will happen at time T" nor that ``\emph{a} will not happen at time T". In other words the \emph{Definition} of future contingents is equivalent to the following

\emph{Principle Q:}
Whether event \emph{a} happens or not at time T depends on information coming from outside space-time at time T.

This \emph{Principle Q} is for me the key feature singling out quantum from classical.

The classical view implies actually that no event is future contingent: Any event can in principle be predicted \emph{with certainty}.

In case of Aristotle's sea battle \emph{Principle Q} amounts to postulate human free will, i.e.: steering of outcomes in human brains happening from outside space-time; this might also be the reason why Thomas Aquinas postulates that ``the universe would not be perfect without randomness" \cite{vs}. Quantum experiments (single-particle Mach-Zehnder interference, single-particle spin 1 Stern-Gerlach, EPR-Bell experiments) allow us to go beyond: We postulate free will on the part of the experimenter and observe correlations between space-like separated detection events; thereby we experimentally demonstrate information coming from outside space-time in devices other than human brains.

\emph{Principle Q} is the core of the standard postulate about the ``collapse of the wavefunction"  Einstein rejected as early as 1927 in the Solvay Congress invoking conflict with relativistic local causality.

Nonlocal correlations that cannot be explained by NOT-faster-than-light communication between detectors is a particular case of \emph{Principle Q}: Which of the possible alternative outcomes happens is unpredictable in principle.

In summary Kochen's ``logical" reconstruction of quantum physics rests on \emph{Principle Q} and is consistent with nonlocal correlations between detection events.

This said, one can't help admiring how much quantum physics was already contained in Aristotle's logical description about the ``sea battle", and Ernst Specker's insight about the relationship between Quantum and ``Infuturabilien".

\section {Can nature be ``more nonlocal" or ``less nonlocal" than quantum?}\label{more-nl}

If one takes violation of ``the CSHS (Bell) inequality" as the basic characteristic of quantum nonlocality then it is tantalizing to ask whether there can be in nature nonlocality beyond the Tsirelson bound. However if one takes ``nonlocality at detection" (as demonstrated in \cite{Guerreiro12}) as axiom of quantum theory then it is obvious that it does not make sense to speak about ``more nonlocality than the quantum one". And the other way around, one cannot have ``less nonlocality than the quantum one": As concluded in \cite{sbsg}, any disappearance of nonlocal coordination between detections violates straightforwardly the conservation of energy in the single quantum events; alternative models assuming ``local parts" would lead to a violation of ``energy conservation" and can be considered refuted by the experiment in \cite{Guerreiro12}.

It is interesting to see that ``Bell inequalities" can be considered a particular case of ``Non-contextuality inequalities", and that assumption of ``exclusivity principle" allows to rule out PR Boxes and other beyond quantum theories \cite{ac08, ac16, abs}. This  confirms the conclusion that ``Bell nonlocality" originates from ``nonlocality at detection", since as stated in Section \ref{nl-context}, ``exclusivity" is another way to formulate nonlocally underpinned conservation of energy. Accordingly, in searching to derive the quantum from principles, it may be convenient to take account of this state of affairs.

As a matter of fact the ``quantum algebra" (leading to violation of Bell inequalities, ``Tsirelson bound" and contextuality) emerges from experiments demonstrating ``single particle interference" (e.g.: Double-slit, Mach-Zehnder): Describing interference requires the notion of ``wave-packet". Shaping a ``wave-packet" means weaving ``locality" (particles) with ``nonlocal threads" (plane-waves) using the tool of ``Fourier transforms". Thus the very introduction of ``wave packets" amounts to assume \emph{local from nonlocal} (paraphrasing Wheeler's ``it from bit") as principle for Quantum, and yields the sinusoidal dependence on phases characteristic of quantum correlations \cite{as10a}. Additionally, ``single particle interference" implies (see Section \ref{nl-context}) that each jointly result of the corresponding detectors should be considered a single measurement result coming from outside space-time. Accordingly different phases define in general different bases and different measurements incompatible with each other: sharp measurements in one basis exclude sharp measurements in a different basis.

On the one hand, the Fourier transform can be generalized first to describe all possible single and multiple qubit systems by means of Hadamard, phase and Pauli operators (see for instance \cite{ae17}), and thereafter to Hilbert spaces with $d\geq3$ through ladder operators. On the other hand the impossibility of sharply measuring both a variable and its Fourier transformed (Heisenberg's uncertainty principle) can be generalized through commutation rules between operators to describe the incompatibility of sharp measurements in different basis.

This means that the sinusoidal dependence on phases leading to the violation of Bell inequalities originates from the Fourier transform. And the same holds for the commutation rules between operators leading to contextuality and the Tsirelson bound. From this perspective ``nonlocality at detection" bears the ``quantum algebra" founding ``contextuality" (characterizing Hilbert spaces with $d \geq 3$) and ``Bell nonlocality" (characterizing Hilbert spaces with $d \geq 4$, $d$ non prime), and rules the whole quantum realm (Hilbert space with $d \geq 2$).

\emph{Local from nonlocal} is the strong physical meaning ``Fourier transforms" contain: Although the founding fathers were not aware of this meaning, it was underpinning all the derivations they achieved to account for the ``wave-particle" behavior appearing in quantum experiments. In any case the founding fathers undoubtedly derived the Quantum from ``a set of experimentally motivated postulates" \cite{cr97}, which in the light of nonlocality and contextuality acquire a much deeper meaning. I think these so improved principles may very well be those we are looking for in seeking to get ``Quantum from principles".

All this makes plain the interest of the ``single-photon space-like Michelson-Morley experiment" \cite{as15} to demonstrate that quantum physics (with the Fourier transform) and relativity (with the Lorentz transformation) have common experimental ground and imply each other. Both quantum nonlocal and relativistic local correlations happen without continuous connection in space-time: Observed events are correlated without ``particles" traveling through trajectories in space-time, that is, without continuous connection in between.

In this respect it is noteworthy that Kochen's reconstruction of quantum mechanics is based on single particle Mach-Zehnder interference experiments and introduces the sinusoidal dependence on phases \cite{fr17}. Even if no physical motivation is given for such a dependence, Kochen's reconstruction makes it plain that any attempt to get Quantum from principles assumes (openly or hiddenly) nonlocal ingredients and the construction of ``wave packets" by means of ``Fourier transforms". And it is also worth mentioning that quantum cryptography (BB84-protocol) and quantum computing (Deutsch-algorithm) have been discovered and first technologically implemented using \emph{single qubit interference}. Thus, although Bell- and contextuality inequalities look as promising resources for developing powerful cryptographic and computing tools, the pioneer quantum technologies reaching the market are based on ``nonlocality at detection".

\section {``Infuturabilien-number" is finite and defines ``laws of nature"}\label{inf-number}

The discussion of the preceding Sections means that the prophet can \emph{contextually} assign values to all possible choices the suitors can do as sketched in Fig.\ref{f3}, \emph{provided his assignment is nonlocal and for all possible experiments}.

The invisible quantum realm is in fact a huge mental ensemble of possible outcomes which is not stored in material memories but rather lies outside space-time (in God's mind), and contains as an essential ingredient all possible setting-choices humans of all times can perform, as sketched in Fig.\ref{f3}. Deutsch's Multiverse \cite{deutsch} is a subset of this ensemble.

By freely choosing the settings of the apparatus here and now the experimenter makes that one of the possible outcome worlds becomes real or visible. The ``freedom" on the part of nature has been some times interpreted in the sense that ``if we have free will, elementary particles must have that ability, too" \cite{ck}, or also ``either no true randomness at all, or omnipresent randomness".\cite{ReWo} In the context of the discussion about``divine omniscience and human free will" this interpretation would correspond to the solution that God neither knows the choice we will make nor its outcome, before we do it.

However the very notion of ``elementary particle's free will" seems self-contradictory for the quantum correlations imply nonlocal spontaneous choices, and `particle' refers to well localized entities. The so called ``spontaneous choices of particles" are nothing other that ``God's spontaneous choices" for assigning outcomes in his mind to all possible human experiments according to distributions that can be predicted by the quantum mechanical ``Born rule". In this sense stating that ``randomness is omnipresent" amounts to state that it results from both human and divine free-will: Randomness is a particular case of free will, and divine omniscience can, in principle, be consistently extended to conditioning on alternatives mutually excluding each other.

This implementation of Specker's parable naturally supports two different interpretations of the Born rule depending on the perspective one takes:

\begin{itemize}
\item As giving the frequency of occurrences of an outcome in the sequence of assignments in the omniscient mind, which is the actual observed frequency in experiment. This interpretation corresponds to the standard one using \emph{``frequentist probabilities"} (called \emph{\textsf{BornF}} in \cite{fr17}).
\item As quantifying a subjective belief of the experimenter about the outcome of a future measurement. This interpretation reflects the incapacity of principle to access the result of a measurement before performing it. It is characteristic of QBism \cite{caf,caf17} and uses \emph{``Bayesian probabilities"}: The probability of an outcome is defined as ``the maximum price a rational agent would pay to enter a bet with payoff 1 if a determined outcome is observed, and no payoff otherwise" (called \emph{\textsf{BornB}} in \cite{fr17})
\end{itemize}

Nonetheless, at the end of the day both interpretations are equivalent, since the assignments in the omniscient mind are done in such a way that the factual frequencies occurring in experiment are well fitted by the Bayesian probabilities.

It is a well known theological theorem that God's infinite power does not mean, he can make a thing to be not made (for this would imply that two contradictories are true at the same time), and likewise he cannot make anything to be absolutely infinite.\cite{sth} Accordingly we assume that the number of all possible measurement choices experimenters of all times can do is very large but finite, and call this the ``Infuturabilien-number" using the term invented by Ernst Specker. Note however that the number of possible mathematical theorems about natural numbers is infinite since these relate to necessary truths, which can be considered participation to God's knowledge by finite minds.

The finiteness of the ``Infuturabilien-number" has noticeable \emph{Implications}:

\begin{enumerate}[{I.}]
\item \emph{The discreteness of space-time}: If space-time were continuum the number of possible setting-choices each human could  in principle do would be uncountable many. The mathematics of continuum Newtonian physics employs to represent the world implies that ``any calculation of complete information about, say, a particle position, would require an infinitely long computation." This fact has profoundly disturbed many physicists \cite{ag03} and still disturbs them today: `` `Real numbers' aren't real at all" (Nicolas Gisin \cite{ng15}).
 \item \emph{Upper bound for signaling} $(v\leq c)$: Without such a bound the number of possible experiments each human being could in principle do even assuming discrete space-time would be countable but infinite.
\item \emph{The heat death of the universe}: Even with the conditions I and II fulfilled, the number of possible choices of experimental settings humans could in principle make would be infinite many, if humanity could live for ever. So the finiteness of the ``Infuturabilien-number" implies that ``within a finite period of time to come the earth must again be, unfit for the habitation of man as at present constituted".\cite{kelvin} This was the way Lord Kelvin formulated the second law of thermodynamics, although obviously one should use the term `universe' instead `earth', if one considers that humanity could in principle live outside the earth.
\end{enumerate}

To get a flavor of the magnitude of the ``Infuturabilien-number" consider again Fig.\ref{f3}. If suitor N had decided to make other choices, he would have produced histories other than the represented one. Assuming 8 rounds and 3 possible choices, the number of possible histories is $3^8$. And the number of possible histories for 10 Suitors would amount to $3^{80}$. We assume signaling bounded by the speed of light and the quantum or pixel of space-time defined by the Planck's length and time. Suppose additionally that the conditions for heat death of the universe lead to death of humanity $10^9$ years from now. Then the number of different settings (space-pixels) a single human being can choose in a given pixel of time amounts roughly to $3 \cdot 10^8 \times 10^9 \times 31\cdot10^6 \times \frac{1}{1.6 \cdot 10^{35}} \approx 10^{60}$ pixels in one direction, and $10^{180}$ space-pixels. Accordingly the number of possible different histories a human being living 100 years ($10^{53}$ time pixels) can generate is of the order $10^{180 \times 10^{53}}= 10^{1.8 \times 10^{55}}$. And the number of possible histories the 7.5 billion people living presently on earth (assuming everyone lives 100 years) could in principle generate is of the order of magnitude of $10^{1.8 \times 10^{55} \times 7.5 \times 10^{9}}\approx 10^{10^{65}}$.

Regarding \emph{Implications I and II}, it is worth mentioning that by reducing gravity to a geometric property of space-time, general relativity unavoidably involves singularities and ``carries within itself the seeds of its own destruction." Thus \emph{Implication I} helps to solve this problem. On the other hand these \emph{Implications} fit to the experiment in \cite{as15}, which (as stated in Section \ref{more-nl}) demonstrates that relativistic local correlations happen without continuous connections. The introduction of the space-time continuum (geometry and ``real" numbers) in physics is a useful idealization, but it should not be considered a ``real" structure underpinning the physical world.\cite{ng15}

Regarding \emph{Implication III}, one could object that Lord Kelvin is not formulating the ``second law" but rather a consequence of it, and that similarly the principle of the finite ``Infuturabilien-number" should be considered a consequence of the second law and not the other way around.

This is an interesting point that brings to light that possibly different principles are subsumed under the term ``second law of thermodynamics", and we discuss this in the next Section \ref{second}.

\section {Death, the ``second law", and the measurement problem} \label{second}

If one assumes Boltzmann's formulation of ``the second law" one is led to infer that ``there is always a non-zero probability ...  of exceptions where the law fails to hold". Boltzmann's thermodynamics does not completely forbid that heat flows from a cold to a hot body, but states only that this happens with exponentially small probability \cite{Wolf17}:

Suppose a gas consisting of \emph{N} molecules fills a room. The molecules could in principle \emph{spontaneously} be all in the left half of the room. This would correspond to an entropy decrease of $\triangle S=N k ln2$ and the probability that this happens is $2^{-N}$.

It is actually astonishing that a ``law" supposed to establish that the ``entropy" (whatever it is) of the universe \emph{always} increases admits that such a quantity may be submitted to ``statistical fluctuations".

To avoid this oddity I would like to propose that the meaning of the ``second law" is not well-defined and actually conflates two different principles:

\begin{itemize}
\item The ``irreversibility" of an event like \emph{the death of humanity}, which can happen only once.
\item The ``irreversibility" of events like \emph{the death of the human person}, which may admit ``exceptions", that is, statistical fluctuations.
\end{itemize}

The ``second law" as formulated by Lord Kelvin relates to \emph{the death of humanity}, and  in this context ``increase of entropy" means \emph{``time arrow"}: the fact that each  pixel of time is a step approaching the ``infuturabilien-number" and diminishing the quantity of remaining choices to perform.

By contrast the ``second law" in the statistical Boltzman formulation relates to processes like \emph{the death of a human person}. Physicians define death as the ``irreversible" break-down of all the brain functions included brainstem. What do they mean by ``irreversible"? Just that a damage happens beyond our capabilities to repair. Another similar process may be ``detection", which in the precedent Sections has been proved to play a key role in Quantum Physics. The process by which an outcome becomes registered (the detection or measurement) is defined with relation to the capabilities of the human observer (the way the human brain functions after all): At detection something happens beyond our capabilities to restore. At detection new information appears in space-time in form of some observable mark or sign (blackening, scintillation, sound). This mark evolves visibly thereafter according to a ``deterministic-looking" pattern, that is, the future of the evolution is determined by the past with probability 1 (actually not \emph{absolutely} 1, as we will see in the coming Sections). The very meaning of quantum measurement is that \emph{when a system collapses, it collapses for all observers} (as stated in Section \ref{copen}).

Accordingly ``Boltzman's second law" should be reformulated to account for the limit of our capacity to repair. A process is \emph{irreversible} because the complexity of the reversing task is beyond our operational capabilities. For the time being the conditions defining this \emph{irreversibility} threshold (likely involving a new constant of nature) are unknown: This is the ``measurement problem". Paraphrasing Stefan Wolf: The key to thermodynamics may lie within the quantum measurement process.\cite{Wolf17}

The ``measurement problem" is somewhat the physical correlate of the ``halting problem" in arithmetics. However in arithmetics we can sharply prove that at any time T there will be questions about numbers that we cannot answer with the methods available at time T (the well known Turing's theorem). By contrast in physics we cannot yet sharply prove (we only intuitively feel) that at any time T there will be physical processes (for instance death) that lie beyond our capabilities to restore, and therefore are irreversible \emph{with relation to the human capabilities}. In this sense interpretations of quantum mechanics can basically be split into two types: those acknowledging ``irreversibility" (with relation to human capabilities) and in particular death as a main constituent of physical reality, and those assuming that humans will once overcome death at will by technical means.

\section {Born's rule and  Schr\"{o}dinger's cat are incompatible}\label{incompat}

As said in the preceding Section \ref{second}, after collapse the registered result follows a deterministic world-line (according to the equations of General Relativity) so that the future can be predicted with \emph{certainty}. Accepting ``superposition of macroscopic objects" (``Schr\"{o}dinger's cat") means to reject ``laws of nature" prescribing definite trajectories, or what is the same, predictions that something will be observed at a certain place with probability 1.

Consequently if one assumes (``Schr\"{o}dinger's cat") one should also reject ``quantum-mechanical laws" imposing that ``the measurement of a quantum system will \emph{with certainty} give a determined outcome".

Frauchiger-Renner's ``assumption (\textsf{Q})" in \cite{fr} includes together ``superposition for arbitrary quantum measurements on cats" \emph{and} ``deterministic statements for quantum measurements on microscopic scales".

To this extent ``assumption (\textsf{Q})" in \cite{fr} seems to be self-contradictory and, if so, the merit of \cite{fr} consists mainly in bringing to light this fact.

So one could think we are contrived to choose between two alternatives regarding the equations we use to describe the physical reality (General Relativity, Born's rule):

\begin{itemize}
\item \emph{They express inexorable laws of nature}, that is, \emph{objective} features intrinsic to physical reality. Accordingly they forbid as well ``Schr\"{o}dinger's cats" as deviations from quantum mechanical probability 1 predictions (like in the case of \emph{certainty} previously mentioned).
\item \emph{They express subjective beliefs (QBism)}. Accordingly they neither exclude ``Schr\"{o}dinger's cats", nor deviations from the quantum mechanical probability assignments.
\end{itemize}

Nonetheless I dare to propose a third alternative blending ``realistic" and ``subjective" aspects: Both, the Born rule and the equations of General Relativity, \emph{ordinarily} hold and can be considered \emph{real} ingredients of our world, nonetheless both admit exceptions under certain circumstances to be specified in the following, and in this sense they are \emph{subjective}. In other words, \emph{there is no inexorable law of nature} that fits \emph{all possible} phenomena that may happen in the world; however the mathematical equations we use to describe and predict the world fit well the \emph{ordinary} regularities relevant to our life and in this sense are \emph{real} ingredients of the world we live in.

I first consider in the next Section \ref{purpose} the case of deviations from the Born rule, and subsequently in Section \ref{cat} the case of
deviations ``deterministic-looking" equations describing the visible (``macroscopic") world, like the equations of General Relativity.

\section {Born's rule, sleep, and purposeful action}\label{purpose}

Consider the outcomes my brain produces while writing this article. I mentally steer my brain's outcomes, that is, the sequences of bits they consist in: I can make that the distribution of bits during a short period deviates from the quantum mechanical predictions for a large number of outcomes. By contrast these could be considered valid for an observer watching me while I am sleeping. The outcome bit-string of a human brain tends towards its most probable sequence, i.e.: a meaningless bit string, in absence of purposeful control (in absence of conscious free-will), that is, during sleep. In presence of purposeful control (awake period), the outcome bit-string human brains ``print" may very well be meaningful, and in fact some times it is!

The Born rule fits well the outcomes of systems acting aimlessly according to a ``random" pattern, and in this sense can be considered to describe \emph{reality} properly. Nonetheless it fails when applied to minds deviating purposefully from random, and hence do not fit the \emph{whole set} of outcomes happening in the world. In this sense the Born rule is ``subjective" or incomplete, in agreement with \emph{QBism}. In particular, the prediction that a measurement upon a determined quantum system (e.g.: a Mach-Zehnder interferometer with equal arm-lengths) yields an outcome with ``probability 1" does not mean that ``nature is obliged" to produce this outcome. In Chris Fuchs' wording: ``All probabilities, including all quantum probabilities, are so subjective they never tell nature what to do. This includes probability-1 assignments."\cite{caf} The use of probability in physics is always an ``idealization" \cite{sk15} ``for all practical purposes".

The Born rule does not \emph{inexorably} apply to the realm of human freedom and creativity.

From this one should not infer we can invoke quantum mechanics to ``explain" so called ``para-normal" phenomena. It is claimed for instance that ``psi-subjects" can change the counting rate of a detector in an interference experiment by acting ``mentally" upon the detector without physically changing the phase-parameters. Such phenomena are not impossible but beyond our technological capabilities: We are as less capable of producing them as putting macroscopic objects in superposition of two distant locations (see next Section \ref{cat}); I cannot act upon devices in the lab the same way as I act upon my brain. Quantum mechanics does not forbid ``para-normal" effects, but it does not provide methods to take hold of them.

\section {Classical ``determinism" and ``Schr\"{o}dinger's cat"} \label{cat}

In the preceding Section \ref{purpose} we have discussed deviations from Born's rule that happen in purposeful action. The other side of the coin are deviations from ``deterministic-looking" patterns in the visible (``macroscopic") world like those described by the equations of Classical Mechanics and General Relativity.

Such equations and rules fit well the outcomes we usually observe. This amazing fact makes it possible for us to behave ``rationally": We would for instance pay no price to enter a bet with payoff 1 if the sun begins to dance in the sky tomorrow at noon, and no payoff if it follows its usual trajectory. In this sense ``deterministic-looking"  rules bearing probability 1 predictions meet a basic content of physical \emph{reality}.

Nonetheless (following QBism) we accept that no probability 1 equation or rule whatsoever can completely fit the \emph{whole set} of outcomes the ``omniscient mind" assigns to any phenomenon. The omniscient mind is kind to us and distributes the \emph{subset} of outcomes we usually observe according to regularities we can grasp, but it may also assign outcomes deviating from these regularities under extraordinary circumstances. The times when these deviations come to happen are unpredictable in principle, the same way as the single outcomes in most quantum experiments are unpredictable. The attempt to cast General Relativity into a probabilistic formulation (Quantum Gravity) supports a QBist perspective: The ``deterministic looking" equations of General Relativity, not less than Born's rule, are utterly ``subjective" or incomplete: \emph{There is no inexorable law of nature.}

If there is no ``law" forbidding nature to produce phenomena which deviate from the regularities we are used to (i.e.: phenomena violating our probability 1 predictions), then (``Schr\"{o}din\-ger's  cats") are not \emph{completely} forbidden. This means that \emph{visible} objects can \emph{in principle} be in a superposition for instance of being at two distant locations: that the sun starts spinning in the sky, or a dead resurrects are not impossible events but only \emph{highly} improbable ones. In other words, they happen with probability 1 in the \emph{ordinary} world we are used to, but not \emph{absolutely}. How low their \emph{absolute probability} is, we don't know, and this data may even be beyond what we will ever know, but in any case the occurrence of such \emph{highly} improbable events does not break any ``inexorable law of nature".

However the reverse implication holds as well: We cannot contrive nature to produce such deviations the same way we contrive it to make a friend's mobile ring by sending him a message. Producing such extraordinary events is beyond our technological capabilities, as it is phoning faster than light.

For this reason we endorse the view that human experimenters cannot cause visible entities (detectors) to be in quantum superposition. By contrast the omniscient mind (the ``prophet" in Specker's parable) containing all the possible worlds can conceive histories with such a superposition, and therefore even produce that observers at a certain place perceive the sun spinning while other observers in another place perceive it as usual (in accord with the result in \cite{fr}, version 1).

When ``irreversible processes" become reversed and phenomena spontaneously deviate from the trajectories that we are used to, people of all times tends to refer to them as ``miracles": In this sense a ``miracle" does not violate any ``law of nature"; it is simply a highly improvable event, sort of ``Schr\"{o}dinger's cat", the realization of which is beyond our \emph{technological} capabilities.

This analysis allows us to complete the conclusion of Section \ref{purpose}:

On the one hand, all rules we use to predict nature do not exhaust the \emph{whole} physical reality (all the outcomes assignments by the ``omniscient mind" and other minds) and in this sense can be considered ``subjective" or incomplete: The Born rule in quantum mechanics and the ``determistic-looking" equations in general relativity are about the decision-making behavior any individual agent should strive for to avoid being Dutch-booked \cite{caf, caf17}.

On the other hand, these rules fit well \emph{a highly significant part} of physical reality: the \emph{subset} of outcomes shaping the regularities we are used to. Thereby they make it possible for us to predict, develop technologies, and live, and in this sense can be considered ``objective" and even an ingredient of the physical reality.

In summary, the set of outcome assignments in the omniscient mind consists of two subsets: one containing the assignments defining the usual regularities we observe, and the other shaping extraordinary phenomena. The equations of General Relativity and the Born rule are at the same time ``objective" (real) and ``subjective" (incomplete).

This also allow us to distinguish two types of visible phenomena: Those we can perform by physical operations (as for instance letting a mobile ring, influencing detection rates by changing phases), and those beyond our physical capabilities (letting a dead resurrect, reversing a detection, changing the counting rate of a detector through ``mental power"). For ``deterministic" and ``random"  phenomena we are able to develop algorithms, but we cannot seize hold of freedom. The distinction between \emph{what is and is not possible} for human experimenters is likely more important than the distinction between \emph{classical and quantum}.

\section {Detectors require rest mass}

In this article we have stressed the central role detectors play in quantum physics: Detectors are devices that make visible information coming from the invisible quantum realm. A non-trivial quantum experiment is always defined \emph{with relation to} a number of detectors, at least 2, and in this sense our view is as relational as Carlo Rovelli's one.\cite{cr97} The set of relevant detectors defines the complete set of projectors and the dimension of the corresponding Hilbert space.

Quantum theory is certainly ``a theory about information".\cite{cr97} However one should specify: ``about information available to the human observer"; ``device independent quantum key distribution" can hardly mean ``distribution without detectors".

Defining accurately what makes a detector to detect, or what is the same establishing when precisely a detection event happens, is part of the unsolved measurement problem. However one can safely say that a detector must be a massif device, \emph{something with rest-mass}, fully in agreement with Bohr's postulate about ``the indispensable use of classical concepts in the interpretation of all proper measurements" \cite{nb}.

In Section \ref{more-nl} we referred to the ``wave packet" to show how local entities (``particles") can be shaped from nonlocal ingredients (``monochromatic waves") using Fourier transforms. This picture would remain incomplete without an explicit reference to the property of \emph{mass}. It is well known that the search for mechanisms explaining how particles (``wave packets") acquire \emph{mass} is currently a main concern in particle physics. In any case \emph{rest mass} is apparently a parameter nature uses to produce entities moving \emph{within} space time, that is, at velocity $v < c $: It is by acquiring mass that local ``energy-packets" become ``matter-packets", the suitable bricks to build the classical visible world. For the aim of this article it is interesting to note the following implication: If the parameter rest mass $m_o$ is defined with relation to energy through $E = m_o c^2$, and one takes $c$ as the upper bound for signaling deriving from \emph{Implication II} in Section \ref{inf-number}, then in the end ``nonlocality at detection" leads us to de Broglie's matter-wave relations and relativity.

Nonetheless this unified picture does not invoke time-ordered causality as ``explanation of the correlations"; here ``Lorentz invariance" or ``covariance" refers rather to the fact that correlations between events are independent of the observer's inertial frame, and are not necessarily tied to time-order \cite{as14}.

\section {``Observers" and ``agents"}

Any science based on observations has necessarily to define the observer who observes. For the science we know and do, observation relates to the human observer. It is a merit of QBism (``participatory realism") to have put the scientist (agent and observer) back into science and stressing that ``reality is more than any third-person perspective can capture".\cite{caf} The measurement problem is tightly related to the role of the observer in quantum physics. The ``quantum of space-time" is defined by the ``minimal distance" two observed events can have. This distance may be given by the Planck scale but for the time being we ignore whether this is the case. To define this ``minimal distance" it is necessary to define sharply what an observation is and when does it happen, that is, to solve ``the measurement problem".

Furthermore one often overlooks that the observer problem affects relativity as well: In special relativity, one invokes the ``grand-father" paradox to exclude signals traveling faster than light, that is, one argues that such signals could be used to kill one's grand-father and this would be absurd. But why is this absurd? Actually because we assume that an observation is something irreversible and one cannot make it to be not happened: Evidence about one's grand-father existence cannot be reversed to non-evidence. But special relativity doesn't tell us the conditions for this irreversibility to emerge, and this means that also this theory has to struggle with the ``measurement problem". In other words, to date both quantum physics and relativity are incomplete.

Interestingly, the singularity problem is related to the observer problem: At any black hole there is a ``world's end" beyond which it does not make sense to speak of observation; the ``world's end" is a hyper-surface defined by either the ``event-horizon" or the ``death-horizon", depending on the size of the black hole.\cite{arsu} Accordingly there is no ``real physical space-time" (there are no ``correlated \emph{visible} things") in the region within the ``world's end", and then there is no ``real physical" black hole singularity either.

Acknowledging the centrality of the observer in quantum mechanics does not mean at all that a human experimenter has to be watching a detector in order this decides to fire or not, but that at detection something happens beyond the human operational capabilities to restore the original state. If this is what the proposal ``quantum mechanics without observer" means, then welcome the proposal. By contrast if it means that ``observation" can be defined without reference to the capabilities of the human observer, the proposal becomes nonsensical: It overlooks the obvious thing that \emph{the human senses are the very foundation of experimental science}.

The difficulty to acknowledge the centrality of the ``human observer" in quantum mechanics (and more in general in science) originates from an epistemological fallacy: One assumes that all what exists does exist in space-time (and therefore is accessible to the human senses), and on the same time one postulates things in space-time that are unaccessible to the human senses!

Consider for instance the predominance of matter over antimatter (\emph{baryon asymmetry}) in the universe we observe today. This imbalance between matter and anti matter is a crucial condition to have stabile bodies and therefore human observers. Baryon asymmetry is supposed to originate from a reaction called ``baryogenesis" at some time in the early universe after inflation stopped. According to the Multiverse ``hypothesis" at this time two parallel universes resulted: the imbalanced one we observe, and a second balanced one. It is obvious that a universe with (\emph{baryon symmetry}) can never be demonstrated on the basis of observations since human observers could not exist in such a universe. So, as already argued in Section \ref{mw}, such a hypothetical alternative does not belong to our physical reality and can only have the status of an idea existing outside space-time in the ``omniscient mind". But this would lead us to an absurdity: By creating space-time at the Big Bang this mind apparently aimed to start a universe capable of being observed by humans, and some micro-seconds later destroyed the very conditions making it possible that observers and observations happen.

This confirms the conclusions in Sections \ref{mw} and \ref{apw}: Any scientific description of the evolution of the universe is necessarily an account of the emergence of human observers. In this sense one can say that physical reality actually starts with the ``arrival" of human observers.

Something similar happens with the role of ``the agent": Quantum contextuality and nonlocality make clear that the free will of the experimenter is a main assumption of the theory, and has as a consequence the ``freedom on the part of nature". But it is often overlooked that the assumption of free-will is crucial in relativity too.\cite{as15}

In summary the Multiverse makes sense only in the form of All-Possible-Worlds, where the alternatives originate from all possible different free choices human agents can do.

Thus Bohr and QBism are both right also regarding the ``observer" and ``agent": The measuring devices (detectors) are excluded from a ``subjective" quantum description when they are heavy enough and capable of obtaining an \emph{irreversible} mark upon them (Bohr, \cite{caf17} p.29). However \emph{``irreversibility"} does not mean ``impossible to reverse absolutely" but impossible to reverse by the human subject (observer and agent). Consequently, the devices are not things separate and foreign to the agent: What happens at detection in the detector is defined by what happens in the human brain while observing; ``the agent and his devices are of one flesh." (\cite{caf17} pp. 29-30).

\section {Conclusion}
None more eloquently than David Deutsch has expressed that any deep statement about the nature of physical reality includes a ``we-perspective":

``The universal quantum computer, in a sense contains within itself all the diversity in nature. No other system does, except perhaps systems that are capable of constructing a universal quantum computer. Certainly we find ourselves unavoidably playing a role at the deepest level of the structure of physical reality." \cite{deutsch}

In my view one will hardly find a better way to bridging Many-Worlds and Copenhagen, and lay the ground-work for the All-Possible-Worlds picture.

The ``hidden assumption" behind ``Many-Worlds" is that the physical reality is defined by ``the free choices human observers can in principle perform". We play a role at the deepest level of the structure of physical reality.

This idea together with ``Infuturabilien" leads to ``All-Possible-Worlds", which is the coherent formulation of ``Many-Worlds": All possible setting-choices humans of all times can perform is a main ingredient of the contextual description of quantum. The other main ingredient is God's contextual outcome assignments to all these possible choices distributed according to ``the \emph{normative} Born rule" (QBism).

The resulting quantum realm is a huge collection of All-Conceivable-Histories, which includes the ``universal quantum computer" as a subsystem. Paraphrasing John A. Wheeler: The entirety of quantum phenomena, rather than being built on particles or fields of force or multidimensional geometry, is built upon billions upon billions of elementary human decisions, ``elementary acts of observer-participancy" (\cite{caf17} p.32, note 35).

Contextuality confirms the result of the experiment in \cite{Guerreiro12}: ``Nonlocality at detection" rules the whole quantum world.
Longing for ``deriving quantum from principles as one derives relativity" may mean that we are overlooking the deep physical meaning behind the Fourier transform: ``Local from nonlocal".

The ``laws of physics" actually arise from the maximal number of experiments the humans of all times can in principle perform: What is and is not possible is not determined by physical ``laws" but the other way around, it is these ``laws" which actually arise from what is and is not possible.

But if the universe only starts with our observations, is then the Big Bang here? To this question John A. Wheeler answered once: ``A lovely way to put it -`Is the big bang here?' I can imagine that we will someday have to answer your question with a `yes'." (\cite{caf}, p. 6, note 5). Without ``human free choices", no physical reality!

\section* {Acknowledgments}

I am thankful to Stefan Wolf and Hugo Zbinden for having invited me to present these ideas in their groups. I acknowledge discussions with the participants to the ``Contextuality Workshop", ETH Z\"{u}rich, Juni 22-23, 2017, in particular Ad\'{a}n Cabello, Joseba Alonso, Marc-Olivier Renou, and Ana Bel\'{e}n Sainz, and comments to the manuscript by Gilles Brassard, Hippolyte Dourdent, Alexei Grinbaum, Chris Fuchs, Simon Kochen, and Renato Renner.

\end{document}